\documentclass[11pt]{article}
\usepackage{amsmath,amssymb,amsfonts,color,graphicx,cite,color,soul}
\usepackage{moriond,epsfig}
\usepackage{axodraw}
\usepackage{slashed}

\def\ee{\end{equation}}
\def\bea{\begin{eqnarray}}
\def\eea{\end{eqnarray}}

\newcommand{\eqFeyn}[1]{%
\begin{array}{c} #1 \end{array}
}
\newcommand{\be}{\beta}
\newcommand{\mD}{m_D}
\newcommand{\mM}{m_M}

\newcommand{\mR}{m_{\tilde R}}

\newcommand{\Anu}{A_\nu}

\newcommand{\Bnu}{B_\nu}
\newcommand{\mnu}{m_\nu}
\newcommand{\Ynu}{Y_\nu}
\def\hSi{\hat{\Sigma}}
\def\Si{\Sigma}

\newcommand{\DRbar}{\ensuremath{\overline{\mathrm{DR}}}}
\newcommand{\mDRbar}{\ensuremath{\mathrm{m}\overline{\mathrm{DR}}}}

\newcommand{\cp}{{\cal CP}}

\newcommand{\edz}{\frac{1}{2}}

\newcommand{\MW}{M_W}
\newcommand{\MZ}{M_Z}

\newcommand{\MA}{M_A}

\newcommand{\Mh}{M_h}
\newcommand{\MH}{M_H}

\newcommand{\Snu}{\tilde \nu}

\newcommand{\tsf}{\theta\kern-.20em_{\tilde{f}}}
\newcommand{\tsfp}{\theta\kern-.20em_{\tilde{f}\prime}}
\newcommand{\tsq}{\theta\kern-.15em_{\tilde{q}}}

\newcommand{\se}[1]{\Sigma_{#1}}
\newcommand{\ser}[1]{\hat{\Sigma}_{#1}}

\newcommand{\KL}{\left(}
\newcommand{\KR}{\right)}

\newcommand{\VL}{\left( \begin{array}{c}}
\newcommand{\VR}{\end{array} \right)}
\newcommand{\ML}{\left( \begin{array}{cc}}
\newcommand{\MLd}{\left( \begin{array}{ccc}}
\newcommand{\MLv}{\left( \begin{array}{cccc}}
\newcommand{\MR}{\end{array} \right)}

\newcommand{\tb}{\tan \beta}

\newcommand{\CTb}{\cot \beta}

\newcommand{\CZb}{\cos 2\beta}

\newcommand{\gev}{\,\, \mathrm{GeV}}
\newcommand{\mev}{\,\, \mathrm{MeV}}

\newcommand{\BC}{\begin{center}}
\newcommand{\EC}{\end{center}}
\newcommand{\BE}{\begin{equation}}
\newcommand{\EE}{\end{equation}}
\newcommand{\BEA}{\begin{eqnarray}}
\newcommand{\EEA}{\end{eqnarray}}
\newcommand{\non}{\nonumber}
\newcommand{\id}{{\rm 1\kern-.12em
\rule{0.3pt}{1.5ex}\raisebox{0.0ex}{\rule{0.1em}{0.3pt}}}}
\newcommand{\lsim}

\newcommand{\gsim}

\newcommand{\tadH}{T_H}
\newcommand{\tadh}{T_h}

\newcommand{\tanb}{\tan \beta\,}

\newcommand{\dmhsq}{\delta m_h^2}

\newcommand{\dZ}[1]{\delta Z_{#1}}

\begin{document}
%\vspace*{2cm}
\begin{flushright}
IFT-UAM/CSIC-11-48
%arXiv:yymm.nnnn [hepph]
\end{flushright}

\title{\Large  \bf $\bf  \Mh$ in MSSM with HEAVY MAJORANA NEUTRINOS}

\author{S.~HEINEMEYER$^{1}$,M.J.~HERRERO$^{2}$, S.~PE\~NARANDA$^{3}$, A.M.~RODR\'IGUEZ-S\'ANCHEZ$^{2}$* \vspace{0.2cm} }

\address{
$^1$Instituto de F\'isica de Cantabria (CSIC-UC), Santander, Spain\\ \vspace{0.1cm}
$^2$Departamento de F\'isica Te\'orica and Instituto de F\'isica Te\'orica,
UAM/CSIC\\
Universidad Aut\'onoma de Madrid, Cantoblanco, Madrid, Spain \\  \vspace{0.1cm}
$^3$Departamento de F\'isica Te\'orica, Universidad de Zaragoza, 
Zaragoza, Spain \\  \vspace{0.1cm}
* Talk given by A.M. Rodriguez-Sanchez at Moriond EW 2011 }

\maketitle\abstracts{ We review the main results  of the one-loop radiative corrections from the 
neutrino/sneutrino sector to the 
lightest Higgs boson mass, $\Mh$, within the context of the so-called 
MSSM-seesaw scenario where right handed neutrinos  and their supersymmetric partners are included in order to
explain neutrino masses.
 For simplicity, we have restricted ourselves to the one generation case. We find sizable corrections to
$\Mh$, which are negative in the region where the Majorana scale is
large ($10^{13}-10^{15} \gev$) 
and the lightest neutrino mass is within a range inspired by data
($0.1-1$~eV). For
some regions of the MSSM-seesaw parameter space, the corrections to $\Mh$
are substantially larger than the anticipated LHC  precision.}

\section*{Introduction}
The current  experimental data 
on neutrino mass differences and 
neutrino mixing angles clearly indicate new physics beyond
the so far successful Standard Model of Particle Physics (SM). In particular, neutrino oscillations imply that at
least two generations of neutrinos must be massive. Therefore, one needs to extend the SM to incorporate neutrino
mass terms.

 We have explored the simplest version of a SUSY extension of the SM, the well known
 Minimal Supersymmetric Standard Model (MSSM), extended by right-handed Majorana neutrinos and
where the seesaw mechanism of type I~\cite{seesaw:I} is implemented to generate the small neutrino masses. 
We focus here in the one generation case. The main advantage of working in 
a SUSY extension of the SM-seesaw is to avoid the huge hierarchy problem induced by the heavy Majorana scale.

On the other hand, it is well known that heavy Majorana neutrinos, with $m_M \sim 10^{13}-10^{15}$ GeV, 
 induce large LFV rates~\cite{LFV}, due to their potentially large Yukawas
to the Higgs sector. For the same reason, radiative corrections to Higgs boson masses due to such heavy Majorana neutrinos
could also be relevant. 
Consequently, our study has been  focused on the 
radiative corrections to the lightest MSSM $\cp$-even $h$ boson mass, $\Mh$,
due to the one-loop contributions from the neutrino/sneutrino sector 
within the MSSM-seesaw framework.

 In the following we briefly review the main relevant aspects of the calculation of the mass corrections and the
numerical results. 
For further details we address the reader to the full version of our work~\cite{Heinemeyer:2010eg},
 where also an extensive list with references to previous works can be found.
%%%%%%%%%%%%%%%%%%%%%%%%%%%%%%%%%%%%%%%%%%%%%%%%%%%%%%%%%%%%%%%%%%%%%%%%%%%%%%%%%%%%%%%%%%%%%%%%%%%%%%
\section*{Calculation}
\subsection*{The neutrino/sneutrino sector}
\label{sec:nusnu}

The MSSM-seesaw model with one neutrino/sneutrino generation is described
in terms of the well known MSSM superpotential plus the new
relevant terms contained in: %~\cite{Gunion:1986yn}
\begin{equation}
\label{W:Hl:def}
W\,=\,\epsilon_{ij}\left[\Ynu \hat H_2^i\, \hat L^j \hat N \,-\, 
Y_l \hat H_1^i\,\hat L^j\, \hat R  \right]\,+\,
\edz\,\hat N \,\mM\,\hat N \,,
\end{equation}
where $\mM$ is the Majorana mass and  $\hat N = (\Snu_R^*, (\nu_R)^c)$ is 
the additional superfield that contains the  right-handed 
neutrino $\nu_{R}$ and its scalar partner $\Snu_{R}$.   

There are also new relevant terms in the soft SUSY breaking potential: 
\begin{equation}
V^{\Snu}_{\rm soft}= m^2_{\tilde L} \Snu_L^* \Snu_L +
  m^2_{\tilde R} \Snu_R^* \Snu_R + (\Ynu \Anu H^2_2 
  \Snu_L \Snu_R^* + \mM \Bnu \Snu_R \Snu_R + {\rm h.c.})~.
\end{equation}

After electro-weak (EW) symmetry breaking, the charged lepton and 
Dirac neutrino masses
can be written as
\begin{equation}
m_l\,=\,Y_l\,\,v_1\,, \quad \quad
\mD\,=\,\Ynu\,v_2\,,
\end{equation}
where $v_i$ are the vacuum expectation values (VEVs) of the neutral Higgs
scalars, with $v_{1(2)}= \,v\,\cos (\sin) \be$ and $v=174 \gev$.

The $ 2 \times 2$ neutrino mass matrix is given in terms of $\mD$ and
$\mM$ by: 
\begin{equation}
\label{seesaw:def}
M^\nu\,=\,\left(
\begin{array}{cc}
0 & \mD \\
\mD & \mM
\end{array} \right)\,. 
\end{equation}
Diagonalization of $M^\nu$ leads to two mass
eigenstates, $n_i \,(i=1,2)$, which are Majorana fermions
with the respective  mass eigenvalues given by:
\begin{equation}
\label{nuigenvalues}
m_{\nu,\, N}  = \edz \KL \mM \mp \sqrt{\mM^2+4 \mD^2} \KR~. 
\end{equation}

In the seesaw limit, i.e. when $\xi \equiv \frac{\mD}{\mM} \ll 1~$
\begin{align}
m_{\nu}= -\mD \xi + \mathcal{O}(\mD \xi^3) \simeq -\frac{\mD^2}{\mM} ~,\quad \quad m_N    &=  \mM + \mathcal{O}(\mD \xi) \simeq \mM ~.
\end{align}
   
Regarding the sneutrino sector, the sneutrino mass matrices for the
$\cp$-even, ${\tilde M}_{+}$,  and the $\cp$-odd, ${\tilde M}_{-}$,
subsectors are given respectively by 
\begin{equation}
{\tilde M}_{\pm}^2=
\left( 
\begin{array}{cc} m_{\tilde{L}}^2 + \mD^2 + \edz \MZ^2 \cos 2 \be & 
\mD (A_{\nu}- \mu \CTb \pm \mM) \\  
\mD (A_{\nu}- \mu \CTb \pm \mM) &
m_{\tilde{R}}^2+\mD^2+\mM^2 \pm 2 \Bnu \mM \end{array} 
\right)~.
\end{equation}
The diagonalization of these two matrices, ${\tilde M}_{\pm}^2$, 
leads to four sneutrino mass eigenstates, ${\tilde n}_i \,(i=1,2,3,4)$. In the seesaw limit,
 where $\mM$ is much bigger than all the
other scales the corresponding sneutrino masses are given by: 
   
\begin{eqnarray}
 m_{{\Snu_+},{\Snu_-}}^2
&=& m_{\tilde{L}}^2 + 
\edz \MZ^2 \CZb \mp 2 \mD (A_{\nu} -\mu \CTb-\Bnu)\xi ~, \non \\
 m_{{\tilde N_+},{\tilde N_-}}^2  &=& \mM^{2} \pm 2 \Bnu \mM + \mR^2 + 2 \mD^2 ~.
\end{eqnarray}

  In the Feynman diagrammatic (FD) approach the higher-order corrected 
$\cp$-even Higgs boson masses in the MSSM, denoted here as $\Mh$ and $\MH$, 
are derived by finding the 
poles of the $(h,H)$-propagator 
matrix, which is equivalent to solving the following equation \cite{mhcMSSMlong}: 
\begin{equation}
\left[p^2 - m_{h}^2 + \hSi_{hh}(p^2) \right]
\left[p^2 - m_{H}^2 + \hSi_{HH}(p^2) \right] -
\left[\hSi_{hH}(p^2)\right]^2 = 0\,.
\label{eq:proppole}
\end{equation}
where $m_{h,H}$ are the tree level masses.
The one loop renormalized self-energies, $\hSi_{\phi\phi}(p^2)$,  in \eqref{eq:proppole} can be expressed
in terms of the bare self-energies, $\Si_{\phi\phi}(p^2)$, the field
renormalization constants $\delta Z_{\phi\phi}$  and the mass counter terms  $\delta m_{\phi}^2$, where $\phi$ stands for
$h,H$.
 For example, the lightest Higgs boson renormalized self energy reads:

\begin{equation}\label{rMSSM:renses_higgssector}
\ser{hh}(p^2)  = \se{hh}(p^2) + \dZ{hh} (p^2-m_h^2) - \dmhsq,
\end{equation}

\subsection*{Renormalization prescription}
 We have used an on-shell renormalization scheme for $\MZ, \MW$ and $\MA$ mass counterterms 
 and $\tadh, \tadH$  tadpole counterterms. On the other hand, we have used a modified $\DRbar$ scheme  $\left(\mDRbar\right)$
 for the renormalization of the wave function 
and $\tanb$. The m$\DRbar$ scheme  is very similar to
the well known $\DRbar$ scheme but instead of subtracting the usual $\Delta= \frac{2}{\epsilon}-\gamma_E+ \log(4 \pi)$ one subtracts
$\Delta_m = \Delta -\log(m^2_M/\mu^2_{\DRbar})$, hence, avoiding large logarithms of the large scale $m_M$.
As studied in other works~\cite{decoup1}, this scheme minimizes higher order corrections
 when two very different scales are involved in a calculation 
of radiative corrections.

\section*{Analytical and Numerical Results}
 In order to understand in simple terms the analytical behavior of our full numerical results we have expanded 
the renormalized self-energies in
powers of the seesaw parameter $\xi=\mD/\mM$:
\begin{equation}
\hSi(p^2)=\left(\hSi(p^2)\right)_{\mD^0}+\left(\hSi(p^2)\right)_{\mD^2}+
\left(\hSi(p^2)\right)_{\mD^4} + \ldots ~.
\label{seesawser}
\end{equation}
The zeroth order of this expansion corresponds to the gauge contribution and it does not depend on $\mD$ or $\mM$. The rest
of the terms of the expansion corresponds to the Yukawa contribution.
The leading term of this Yukawa contribution is the ${\cal O}(m_D^2)$ term, because 
it is the only one not suppressed by the Majorana scale. 
In fact it goes as $Y_\nu^2M^2_{\rm EW}$, where  $M^2_{\rm EW}$ denotes generically the electroweak scales involved, 
concretely, $p^2$, $M_Z^2$ and $M_A^2$. In particular, the ${\cal O}(p^2m_D^2)$  terms of the renormalized self-energy,
 which turn out to be among the most relevant leading contributions, separated into the neutrino and sneutrino contributions,
 are the following: 
\vspace{0.3cm}
%%%%%%%%%%%%%%%%%%%%%%%%%%%%%
\begin{equation}
\left.\hSi_{hh}^{\mDRbar}\right|_{\mD^2 p^2} \sim
\left.
\eqFeyn{%
\begin{picture}(80,50)
\DashLine(10,25)(25,25){3}
\Text(8,25)[r]{$h$}
\DashLine(55,25)(70,25){3}
\Text(72,25)[l]{$h$}
\ArrowArc(40,25)(15,0,180)
\Text(40,46)[b]{$\nu_L$}
\ArrowArc(40,25)(15,-180,0)
\Text(40,4)[t]{$\nu_R$}
\end{picture}
}
+
\eqFeyn{%
\begin{picture}(80,50)
\DashLine(10,25)(25,25){3}
\Text(8,25)[r]{$h$}
\DashLine(55,25)(70,25){3}
\Text(72,25)[l]{$h$}
\DashArrowArc(40,25)(15,0,180){3}
\Text(40,46)[b]{$\Snu_L$}
\DashArrowArc(40,25)(15,-180,0){3}
\Text(40,4)[t]{$\Snu_R$}
\end{picture}
}\right|_{\mD^2 p^2}
\sim \hspace{0.3cm}\frac{g^2 p^2 \mD^2 c^2_{\alpha}}{64 \pi^2  \MW^2 s^2_{\beta} } + \frac{g^2 p^2 \mD^2 c^2_{\alpha}}{64 \pi^2  \MW^2 s^2_{\beta} }
\vspace{0.3cm}
\end{equation}
Notice that the above neutrino contributions come from the Yukawa interaction
 $g_{h\nu_L \nu_R} =  -\frac{igm_D \cos\alpha}{2M_W\sin\beta}$, which is extremely suppressed in the Dirac case but can
be large in the present Majorana case.
 On the other hand, the above sneutrino contributions come from
the new couplings 
$g'_{h\Snu_L\Snu_R}=-\frac{igm_Dm_M\cos\alpha}{2M_W\sin\beta}$, which are not
present in the Dirac case. 
It is also interesting to remark  that these terms, being $\sim p^2$ are absent in both the effective potential  and the RGE approaches.

\begin{figure}[ht!]
   \begin{center} 
     \begin{tabular}{c} \hspace*{-12mm}
  	\psfig{file=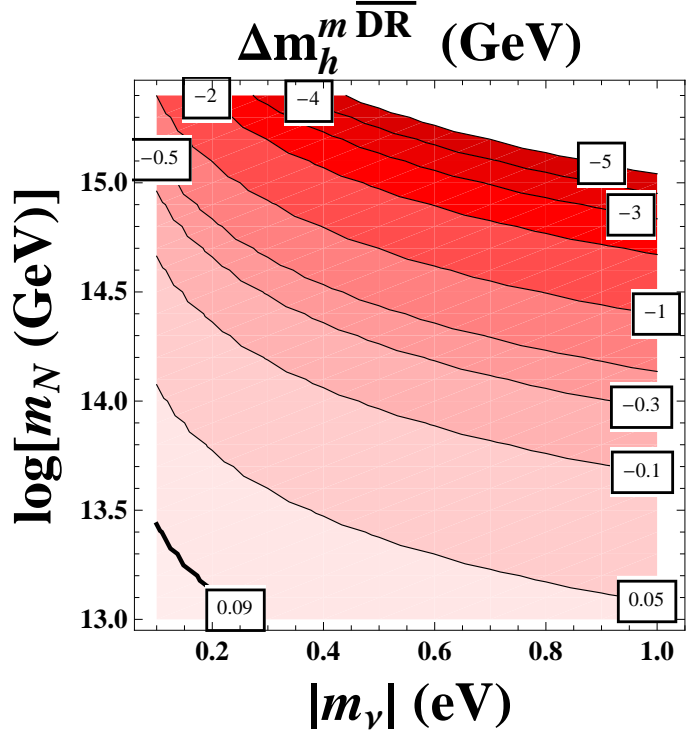,width=90mm,clip=}   
       \end{tabular}
     \caption{Contour-lines for the Higgs mass corrections from the
     neutrino/sneutrino sector as a function of the physical  
     Majorana neutrino masses, light $|\mnu|$ and heavy $m_N$. The other parameters are 
     fixed to: $A_\nu=\Bnu=m_{\tilde L}=\mR=
     10^3 \gev$, $\tb=5$, $M_A=200 \gev$ and $\mu=200 \gev$.}  
   \label{masscontours1} 
   \end{center}
 \end{figure}

With respect to the numerical results, figure  \ref{masscontours1} exemplifies the main features of the extra Higgs mass corrections 
$\Delta m_h^{\mDRbar}$
due to neutrinos and sneutrino loops
 in terms of the two physical Majorana neutrino masses, $m_N$ and $\mnu$.
 For values of $m_N < 3 \times 10^{13}$ GeV and $|\mnu|< 0.1 -0.3$~eV
 the corrections to $M_h$ are positive and smaller than 0.1 GeV. In this region, the gauge contribution dominates. In fact,
 the wider black  contour line with fixed $\Delta m_h^{\mDRbar}=0.09$
coincides with the prediction for the case where just the gauge part in the
self-energies have been included. This means that 'the distance' of any other
contour-line respect to this one represents the difference in the
radiative corrections respect to the MSSM prediction.

However, for larger values of  $m_N$ and/or $|\mnu|$ the Yukawa part dominates, and the radiative corrections
become negative and larger in absolute value, up to  values of -5 GeV in the right upper
 corner of Fig \ref{masscontours1}. These corrections grow in modulus  proportionally to $\mM$ and $\mnu$, due to the fact
that the seesaw mechanism impose a relation between the three masses  involved,
$\mD^2 = |m_{\nu}| m_N$.
%%%%%%%%%%%%%%%%%%%%%%%%%% F I G U R E %%%%%%%%%%%%%%%%%%%%%%%%%%%%%%%%%%%% 
%%%%%%%%%%%%%%%%%%%%%%%%%%%%%%%%%%%%%%%%%%%%%%%%%%%%%%%%%%%%%%%%%%%%%%%%%%%%%%%%%%%%%%%%%%%

\newpage
%%%%%%%%%%%%%%%%%%%%%%%%%%%%%
\section*{Conclusions}
We have used the Feynman diagrammatic approach for the calculation of the radiative corrections to
the lightest Higgs boson mass of the MSSM-seesaw.
This method  does not neglect the external momentum of the incoming and outgoing particles as it happens 
in the effective potential approach. We have performed a full calculation, obtaining not only the leading logarithmic terms 
as it would be the case in a RGE computation but also the finite terms, that
we have seen that can be sizable for heavy Majorana neutrinos ($10^{13}-10^{15} \gev$) and the lightest neutrino mass
within a range inspired by data
($0.1-1$~eV). For some regions of the MSSM-seesaw parameter space, the corrections to $\Mh$
are substantially larger ( up to -5 GeV) 
than the anticipated LHC precision ($\sim 200 \mev$)~\cite{lhctdrs}.

\end{document}